\documentclass[]{spie}  

\usepackage{amsmath,amsfonts,amssymb}
\usepackage[dvipdfmx]{graphicx}
\usepackage[dvipdfmx]{color}
\usepackage[colorlinks=true, allcolors=blue]{hyperref}

\title{Feasibility study of upper atmosphere density measurement on the ISS by observations of the CXB transmitted through the Earth rim}

\author[a]{Takumi Kishimoto}
\author[a]{Kumiko K. Nobukawa}
\author[b]{Ayaki Takeda}
\author[c]{Takeshi G. Tsuru}
\author[d]{Satoru Katsuda}
\author[e]{Nakazawa Kazuhiro}
\author[b]{Koji Mori}
\author[f]{Masayoshi Nobukawa}
\author[c]{Hiroyuki Uchida}
\author[a]{Yoshihisa Kawabe}
\author[a]{Satoru Kuwano}
\author[b]{Eisuke Kurogi}
\author[a]{Yamato Ito}
\author[a]{Yuma Aoki}
\affil[a]{Faculty of Science and Engineering, Kindai University, 3-4-1 Kowakae, Higashi-Osaka 577-8502, Japan}
\affil[b]{Department of Applied Physics and Electronic Engineering, Faculty of Engineering, University of Miyazaki,1-1 Gakuen Kibanadai Nishi, Miyazaki, Miyazaki 889-2192, Japan}
\affil[c]{Department of Physics, Faculty of Science, Kyoto University, Kitashirakawa Oiwake-cho, Sakyo-ku, Kyoto 606-8502, Japan}
\affil[d]{Graduate School of Science and Engineering, Saitama University, 255 Shimo-Ohkubo, Sakura, Saitama 338-8570, Japan}
\affil[e]{Kobayashi-Maskawa Institute for the Origin of Particles and the Universe, Nagoya University, Furo-cho, Chikusa-ku, Nagoya, Aichi 464-8601, Japan}
\affil[f]{Faculty of Education, Nara University of Education, Takabatake-cho, Nara, Nara 630-8528, Japan}

\authorinfo{Further author information: (Send correspondence to T.K.)\\T.K.: E-mail: kishimoto.takumi@kindai.ac.jp}

\begin{document} 
\maketitle

\begin{abstract}
Measurements of the upper atmosphere at $\sim100$~km are important to investigate climate change, space weather forecasting, and the interaction between the Sun and the Earth. Atmospheric occultations of cosmic X-ray sources are an effective technique to measure the neutral density in the upper atmosphere. We are developing the instrument SUIM dedicated to continuous observations of atmospheric occultations. SUIM will be mounted on a platform on the exterior of the International Space Station for six months and pointed at the Earth's rim to observe atmospheric absorption of the cosmic X-ray background (CXB).  
In this paper, we conducted a feasibility study of SUIM by estimating the CXB statistics and the fraction of the non-X-ray background (NXB) in the observed data. 
The estimated CXB statistics are enough to evaluate the atmospheric absorption of CXB for every 15~km of altitude.
On the other hand, the NXB will be dominant in the X-ray spectra of SUIM. Assuming that the NXB per detection area of SUIM is comparable to that of the soft X-ray Imager onboard Hitomi, the NXB level will be much higher than the CXB one and account for $\sim80$\% of the total SUIM spectra. 
\end{abstract}


\section{Introduction}
\label{sec:intro}  
Studying the upper atmosphere at the altitude of $\sim100$~km is important from the perspectives of climate change, space weather forecasting, and the interaction between the Sun and the Earth [\citenum{Roble1989}, \citenum{Marubashi1989}]. On the other hand, the data of the upper atmosphere around this altitude has been very limited because it is difficult to observe it by in-situ instruments like balloons or satellites. Determan et al. (2007) [\citenum{Determan2007}] demonstrated that atmospheric occultations of cosmic X-ray sources are a unique technique to measure the neutral density in the upper atmosphere by analyzing the data of the Crab Nebula and Cygnus X-2.  
 Katsuda et al. (2021, 2023) [\citenum{Katsuda2021}, \citenum{Katsuda2023}] measured the vertical density profile of the Mesosphere and Lower Thermosphere  by systematical analysis of the data during the atmospheric occultations of the Crab Nebula observed by X-ray astronomy satellites. During atmospheric occultations, X-ray spectra alter due to photo-absorption by the atmosphere, and spectral-fitting derives atmospheric densities. However, the data obtained by X-ray astronomy satellites were sparse because X-ray astronomy satellites can observe atmospheric occultations for only short duration (typically 1 minutes every orbit) just before/after occultations of the solid Earth.

We are developing a new instrument dedicated to continuous observations of atmospheric occultations, Soipix for observing Upper atmosphere as Iss experiment Mission (SUIM).  This instrument is planned to be mounted on the International Space Station for six months in 2025, utilizing a platform on the exterior of the ISS, the Materials International Space Station Experiment (MISSE). 
The X-ray detectors onboard SUIM are our developing new pixel imaging sensors, SOIPIX, which are based on Silicon-On-Insulator (SOI) CMOS technology. 
SUIM has a collimator placed in front of the detectors to observe atmospheric absorption of the cosmic X-ray background (CXB) in the energy range of 3--10~keV, with determinating incident angles to measure atmospheric neutral density at each altitude in the range of 60--150~km. Figure~1 represents a schematic view of the SUIM observations. 
In this paper, we conduct a feasibility study of SUIM by estimating the CXB statistics that would be obtained during the six-month observation and also by predicting the fraction of Non X-ray background (NXB) in the observation data.

   \begin{figure} [ht]
   \begin{center}
   \begin{tabular}{c} 
   \includegraphics[height=4.5cm]{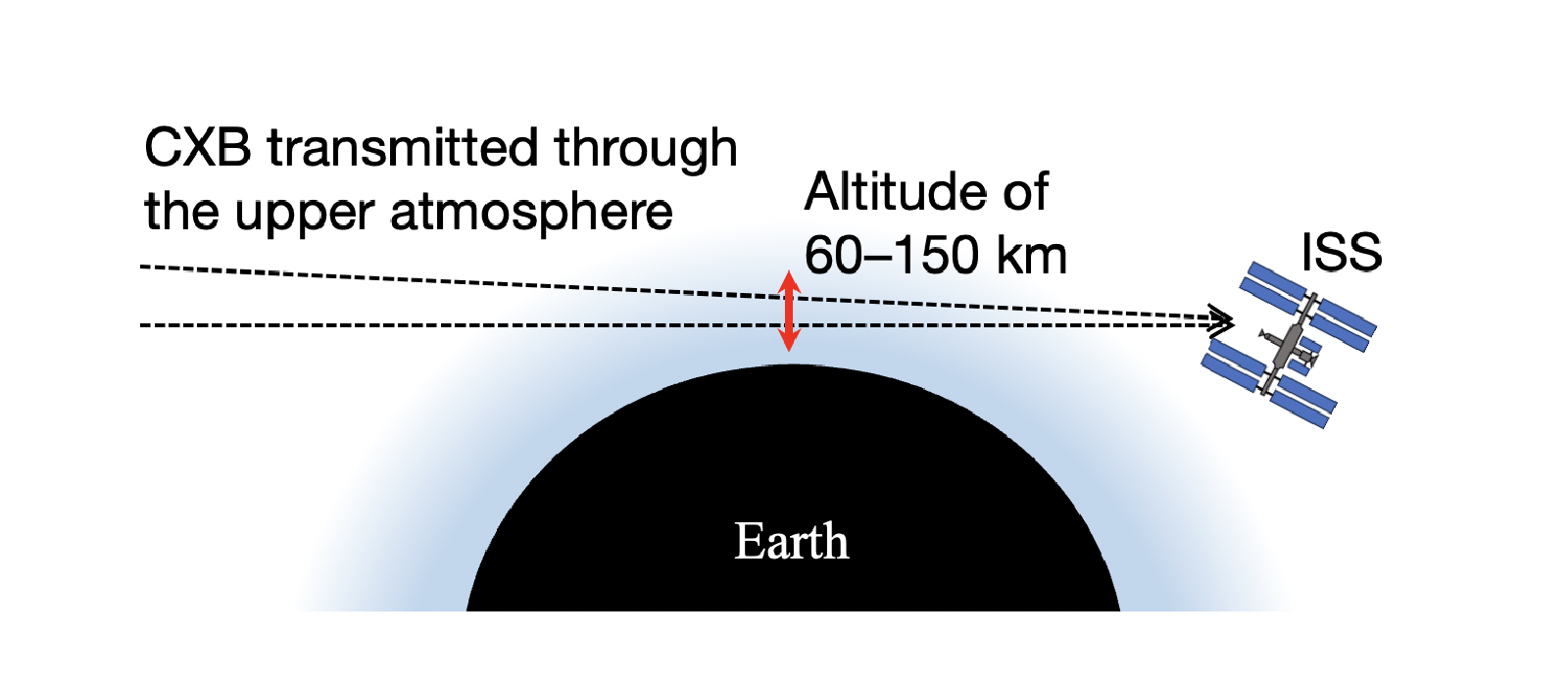}
	\end{tabular}
	\end{center}
   \caption[example] 
   { \label{fig:example} 
   Schematic view of atmospheric observations by SUIM. 
   }
   \end{figure}

\section{Estimation of CXB photons statistics}
Since the CXB is a diffuse emission, its statistics are proportional to the product of three parameters: the field of view (FOV), detection area, and effective observation time. In the following, we describe the three parameters in SUIM and then estimate the CXB statistics. 

\subsection{FOV and detection area of SUIM}
\label{sec:title}
The FOV and detection area are constrained by the detector configuration. Figure~2 shows the detector configuration and table 1 summarizes the parameters. 
SUIM will be equipped with the collimator with sixteen slits, eight for each X-ray detector. The slits will be aligned parallel to the Earth's horizon. Every 73 (vertical) $\times$ 300 (horizontal) pixels on each detector will observe the atmosphere from each slit. The pixel size is $36~\mu$m$\times36~\mu$m, and the FOV per each pixel is 0.33~deg (vertical) $\times$ 22.0~deg (horizontal). The vertical FOV corresponds to an altitude range of $\sim15$~km. 

\begin{figure} [ht]
   \begin{center}
   \begin{tabular}{c} 
   \includegraphics[height=6cm]{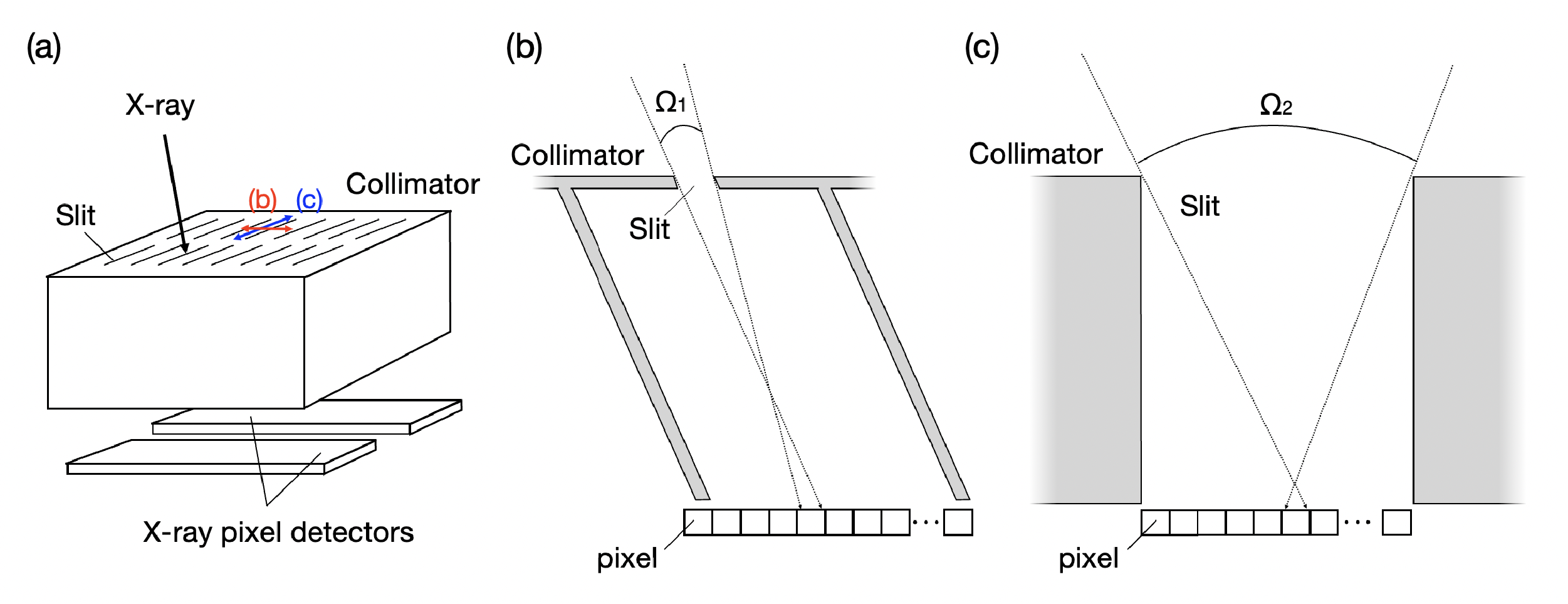}
	\end{tabular}
	\end{center}
   \caption[example] 
   { \label{fig:example} 
   (a) Geometry of the collimator and two X-ray pixel detectors in SUIM. (b) Cross-sectional view perpendicular to a slit. (c) Cross-sectional view along a slit. 
   }
   \end{figure}

\begin{table}[ht]
\caption{Parameters of the SUIM detector} 
\label{tab:fonts}
\begin{center}

\begin{tabular}{ll} 
\hline
\rule[-1ex]{0pt}{3.5ex}   Parameters & Values  \\
\hline
\hline
\multicolumn{2}{l}{\rule[-1ex]{0pt}{3.5ex}FOV per pixel (spatial resolution)} \\
\rule[-1ex]{0pt}{3.5ex} ~~~~~~$\Omega_{1}$ (vertical) & 0.33~deg (corresponds to $\sim15$~km) \\
\rule[-1ex]{0pt}{3.5ex}  ~~~~~~$\Omega_{2}$ (horizontal) & 22.0~deg   \\
\rule[-1ex]{0pt}{3.5ex}  Detection area per pixel & $36~\mu$m$\times36~\mu$m \\
\rule[-1ex]{0pt}{3.5ex} Number of pixels per slit & 73 (vertical) $\times$ 300 (horizontal)    \\
\rule[-1ex]{0pt}{3.5ex} Number of slits per X-ray detector & 8   \\
\rule[-1ex]{0pt}{3.5ex}  Number of X-ray detectors & 2 \\
\hline
\end{tabular}
\end{center}
\end{table} 

\subsection{Effective observation time of SUIM}
The effective observation time is constrained by two factors: (1) the background count rate and (2) the fluctuation of the ISS attitude. 
The X-ray pixel detectors SOIPIX are sensitive to charged particles as well as X-rays and have a dead time of 400~$\mu$s at each incidence.
It indicates that the higher the background count rate, the lower the detection efficiency. We simulated the detection efficiency of the CXB assuming the background count rate in the ISS orbit, based on the data of the Radiation Belt Monitor (RBM)\footnote{\url{https://data.darts.isas.jaxa.jp/pub/maxi/rbm/}}, which is equipped with the Monitor of All-sky X-ray Image (MAXI) on the ISS. Figure~3 indicates a histogram of the charged-particle count rate obtained by the MAXI/RBM for six months (gray). Although the mode is $\sim1$~counts~s$^{-1}$, but the background count rate can sometimes exceed $10^4$~counts~s$^{-1}$. We note that the charged particle count rate generally has a negative correlation with the Cut-off Rigidity (COR), but when the ISS passes near the poles, the charged-particle count rate can increase very rapidly due to Relativistic Electron Precipitation [\citenum{Ueno2019}].  We simulated the detection efficiency of the CXB at each histogram bin and obtained the averaged observation efficiency in the ISS orbit to be $\sim89$\%.

Since the attitude of the ISS is constantly fluctuating, the target altitude (verti) can be out of the FOV. We calculated the percentage of time in a day that the target altitude is in the FOV based on the actual ISS attitude data as represented by figure~4. We estimated that the target altitude is in the FOV for $\sim81$\% of the observation. 

In addition to the above two factors, we also assume that SUIM may observe only at night. This is because MISSE has a resettable fuse that reduces power at high temperatures, and the temperatures may exceed acceptable limits during the day depending on the beta angle, which can makes observations unavailable.
Taking into account all of the above, we estimate the effective observation time to be $5.7\times10^6$~s for a six-month exposure ($\sim36$\% of the total exposure time).

 \begin{figure} [ht]
    \begin{minipage}[t]{0.45\textwidth}
    \centering
   \includegraphics[height=5cm]{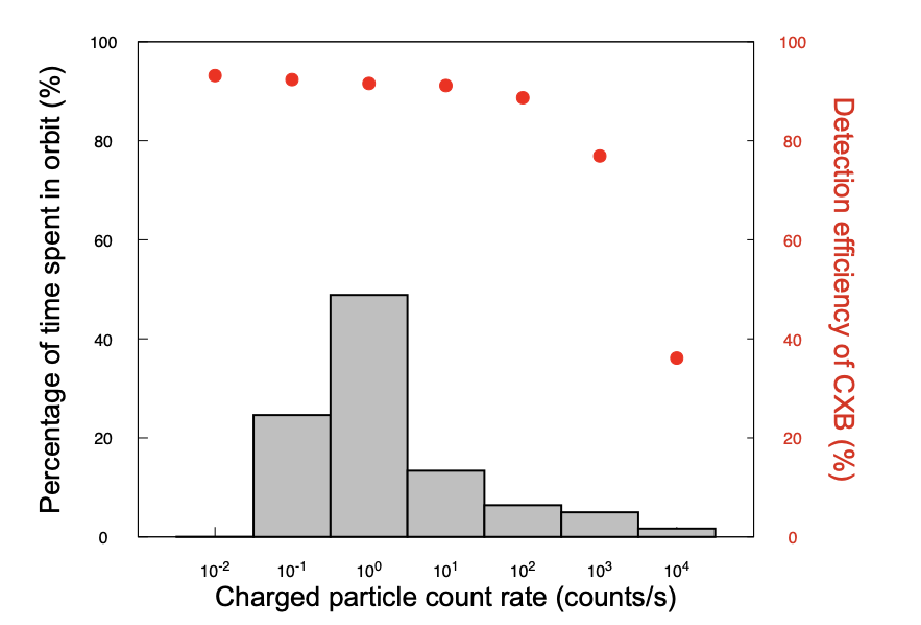}
   \caption { \label{fig:example} 
   (gray) Histogram of charged-particle count rate in the ISS orbit. (red) Simulated detection efficiency of the CXB at each bin. 
   }
   \end{minipage}
   \hspace{0.04\columnwidth}
    \begin{minipage}[t]{0.5\textwidth}
    \centering
   \includegraphics[height=5cm]{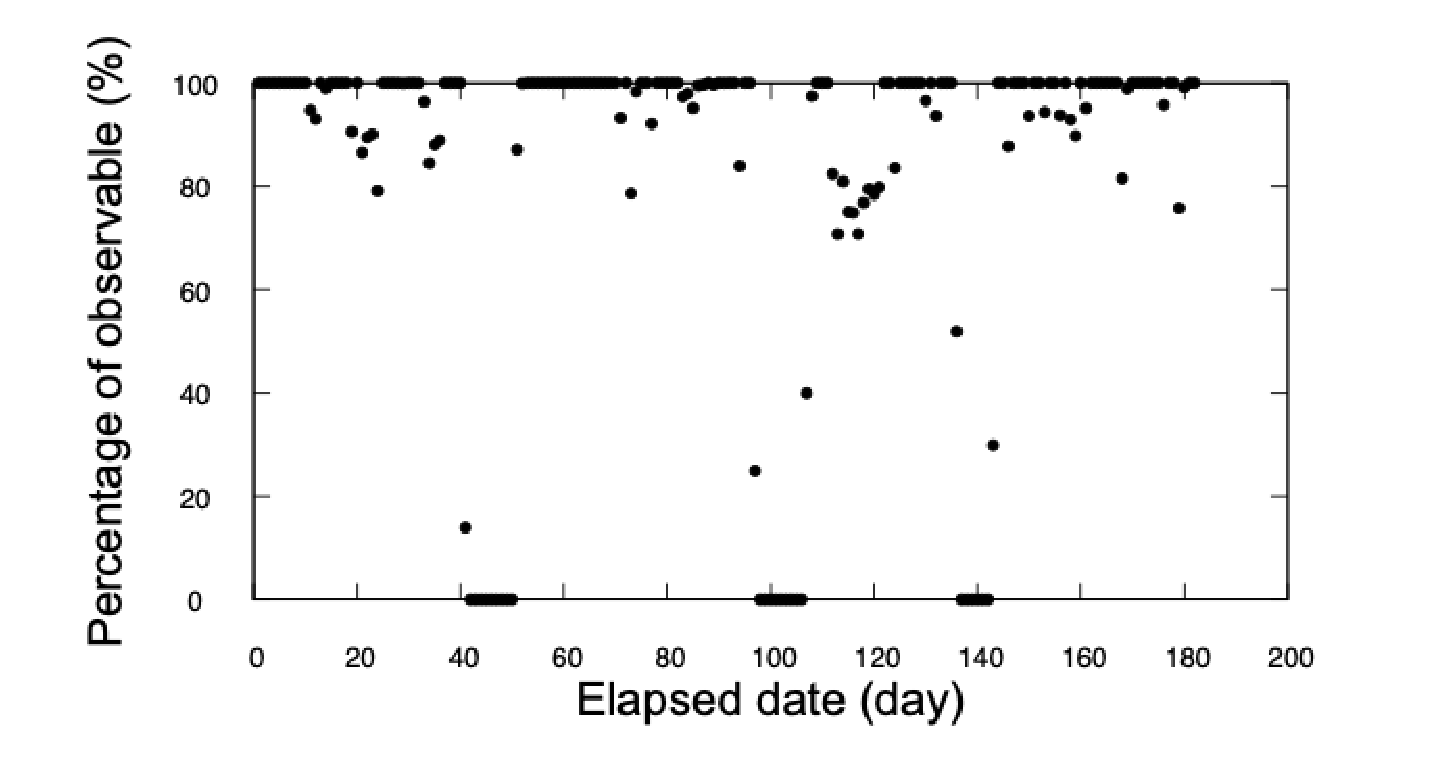}
   \caption{ \label{fig:example} 
   Percentage of time in a day that the target altitude is in the FOV assuming the actual ISS attitude data.
   }
   \end{minipage}
   \end{figure} 
   
\subsection{CXB photon statistics in six months}
We adopt the CXB flux in the 3--10 keV band of $1.82\times10^{-3}$~ photon~s$^{-1}$~cm$^{-2}$~deg$^{-2}$ according to Kushino et al (2002) [\citenum{Kushino2002}].
Based on the FOV, detection area, and effective observation time of SUIM, we expect one photon per pixel to be detected in six months.
It indicates that $4.8\times10^3$~photons are expected to be obtained for every 15~km of altitude, which is enough statistics to evaluate the atmospheric absorption of the CXB.

\section{Estimation of the NXB fraction}
An accurate estimation of the NXB in SUIM should require a particle simulation, but here we assume that the NXB per detection area of SUIM is comparable to that of the soft X-ray Imager (SXI) [\citenum{Tanaka2018}] onboard Hitomi [\citenum{Takahashi2018}], and compare the CXB level expected in SUIM with the NXB. Hitomi is an X-ray astronomy satellite in the low Earth orbit with an altitude of 575~km, and the SXI is an X-ray CCD camera. The NXB level of the SXI depends on COR [\citenum{Nakajima2018}]. Figure~5 shows the NXB spectra of the SXI scaled by the detection area of SUIM, compared with the CXB level expected in SUIM. The CXB level will be comparable to that of the NXB at $\sim3$~keV, but the NXB will account for $\sim80$\% of the total SUIM spectrum above 3~keV. Hence it would be necessary to subtract estimated NXB spectra from the total ones to obtain atmospheric absorbed CXB spectra. The SUIM team is considering that, for example, a portion ($\sim15$\%) of the detector could be shaded to measure only the NXB spectra.

   \begin{figure} [ht]
   \begin{center}
   \begin{tabular}{c} 
   \includegraphics[height=5cm]{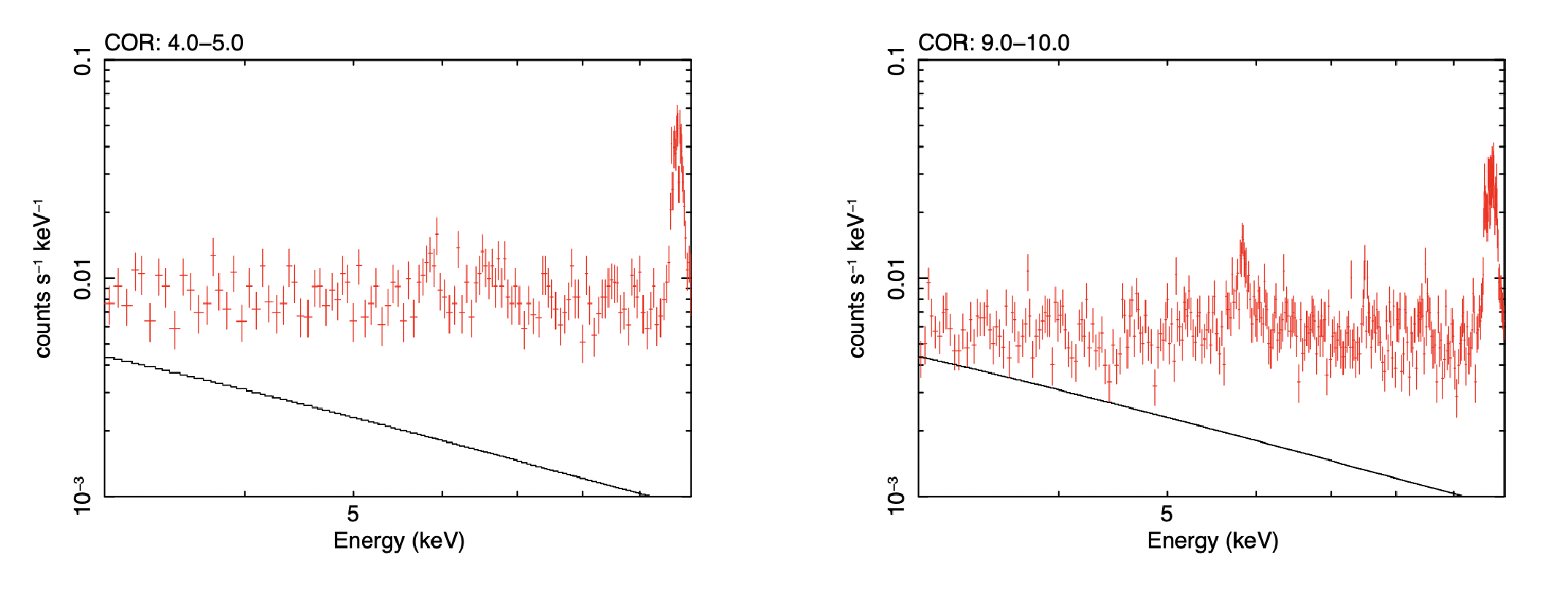}
	\end{tabular}
	\end{center}
   \caption[example] 
   { \label{fig:video-example} 
(red) NXB spectrum obtained by Hitomi/SXI \cite{Nakajima2018} at the COR of 4.0--5.0~GV (left) and 9.0--10.0~GV (right), scaled by the detection area of SUIM. (black) CXB level expected in SUIM.}
   \end{figure}

\section{Conclusion}
We conducted the feasibility study of SUIM by estimating the CXB statistics and the fraction of NXB in the observed data. Based on the FOV, detection area, and effective observation time of SUIM, the CXB statistics during the six-month observation are estimated to be one photon per pixel, which would be enough statistics to evaluate the atmospheric absorption of the CXB ($4.8\times10^3$~photons for every 15~km of altitude). The problem will be the NXB level. Assuming that the NXB per detection area of SUIM is comparable to that of Hitomi/SXI, we predict that the NXB will be dominant ($\sim80$\%) in the SUIM spectra above 3~keV, far beyond the CXB level. It would be necessary to take measures such as shading a portion ($\sim15$\%) of the detector to measure only the NXB spectra and subtracting them from the observed total spectra.

\acknowledgments 
 We are grateful to Satoshi Nakahira, Ken Ebisawa, and Mikio Morii for helpful advice on the MAXI data analysis and the background in the ISS orbit. This work was supported by Grants-in-Aid for Scientific Research from the Ministry of Education, Culture, Sports, Science and Technology (MEXT) of Japan, No. 23H00151, and MEXT Coordination Funds for Promoting Aerospace Utilization, Japan Grant Number JPJ000959. This work was also achieved by Murata Science and Education Foundation, and Mitsubishi Foundation.

\bibliography{report} 
\bibliographystyle{spiebib} 

\end{document}